\documentclass[aps,twocolumn,showpacs,prl]{revtex4}
\usepackage{graphicx}
\begin{document}
\title{Random Walks on Complex Networks}
\author{Jae Dong Noh}
\affiliation{Department of Physics, Chungnam National University, Daejeon
305-764, Korea}
\author{Heiko Rieger}
\affiliation{Theoretische Physik, Universit\"at des Saarlandes, 66041
Saarbr\"ucken, Germany}
\date{\today}

\begin{abstract}
We investigate random walks on complex networks and derive an
exact expression for the mean first passage time (MFPT) between two nodes.
We introduce for each node the random walk centrality $C$, which is the 
ratio between its coordination number and a characteristic relaxation time, 
and show that it determines essentially the MFPT. The centrality of a node
determines the relative speed by which a node can receive and 
spread information over the network in a random process. Numerical
simulations of an ensemble of random walkers moving on paradigmatic
network models confirm this analytical prediction.
\end{abstract}
\maketitle

From biology over computer science to sociology the world is abundant
in networks.  Watts and Strogatz~\cite{Watts.Strogatz98} demonstrated
that many of these real-world networks pose characteristic features
like the small world and clustering property. Since classical network
models do not display these features, new models have been developed
to understand the structure underlying real-world networks
~\cite{Albert.Barabasi02,Dorogovtsev.Mendes02}. The important finding
is that there exists a scale-free~(SF) network characterized by a
power-law degree distribution~\cite{Albert.Jeong.Barabasi99}, that is
also a characteristic for the World-Wide-Web~(WWW)
\cite{Albert.Jeong.Barabasi99}, and for many other networks in various
disciplines~\cite{Albert.Barabasi02}.

The SF network has a heterogeneous structure. For instance, the WWW
analyzed in Ref.~\cite{Albert.Jeong.Barabasi99} includes nodes with
degrees ranging from ${\cal O}(1)$ to ${\cal O}(10^3)$. The
heterogeneity leads to intriguing properties of SF networks. In the
context of percolation, SF networks are stable against random removal
of nodes while they are fragile under intentional attacks targeting on
nodes with high degree~\cite{Cohen.E.A.H00,Albert.Jeong.Barabasi00}.
Statistical mechanical systems on SF networks also display interesting
phase transitions~\cite{Pastor-Satorras.Vespignani01,
Dorogovtsev.Goltsev.Mendes02,Igloi.Turban02}. In a transport process,
each node in SF networks does not contribute equally likely.  The
importance of each node in such a process is measured with the
betweenness centrality~\cite{Newman01}, which has a broad power-law
distribution~\cite{Goh01.02}.

In this paper, we study a random walk on general networks with a particular
attention to SF networks. The random walk is a fundamental dynamic 
process~\cite{Hughes95}. It is theoretically interesting to study 
how the structural heterogeneity affects the nature of the diffusive 
and relaxation dynamics of the random walk~\cite{RWref}.
Those issues will be studied further elsewhere~\cite{Noh.Rieger03}.
The random walk is also interesting since it could be a mechanism of 
transport and search on networks~\cite{Adamic.L.P.H01,Guimera.etal02,Holme03}.
Those processes would be optimal if one follows the shortest path
between two nodes under considerations. 
Among all paths connecting two nodes, the shortest path is given by the one 
with the smallest number of links~\cite{comment1}.  
However the shortest path can be found only after
global connectivity is known at each node, which is improbable 
in practice. The random walk becomes important in the extreme opposite case 
where only local connectivity is known at each node.
We also suggest that the random walk is a useful tool in studying the
structure of networks. 

In the context of transport and search, the mean first passage time~(MFPT)
is an important characteristic of the random walk.
We will derive an exact formula for the MFPT of a random walker from
one node $i$ to another node $j$, which will be denoted by $\langle
T_{ij}\rangle$, in arbitrary networks.
In the optimal process it is just given by the number of links in the
shortest path between two nodes, and both motions to one direction
and to the other direction are symmetric.
However, a random walk motion
from $i$ to $j$ is not symmetric with the motion in the opposite direction.
The asymmetry is characterized with the difference in the MFPT's.
It is revealed that the difference is determined by a potential-like quantity
which will be called the {\em random walk centrality}~(RWC).
The RWC links the structural heterogeneity to the asymmetry in dynamics.
It also describes centralization of information wandering over networks.

We consider an arbitrary {\em finite} network (or graph) which consists
of nodes $i=1,\ldots,N$ and links connecting them. We assume that the
network is connected (i.e. there is a path between each pair of nodes
$(i,j)$), otherwise we simply consider each component separately. The
connectivity is represented by the adjacency matrix ${\bf A}$ whose
element $A_{ij} = 1~(0)$ if there is a link from $i$ to $j$~(we set
$A_{ii}=0$ conventionally). In the present work, we restrict ourselves
to an {\em undirected} network, namely $A_{ij} = A_{ji}$.  The degree,
the number of connected neighbors, of a node $i$ is denoted by $K_i$
and given by $K_i = \sum_{j} A_{ij}$.

The stochastic process in discrete time that we study is a 
random walk on this network described by a master equation. The
transition probabilities are defined by the following rule: A walker
at node $i$ and time $t$ selects one of its $K_i$ neighbors with {\it
equal} probability to which it hops at time $t+1$, thus the transition
probability from node $i$ to node $j$ is $A_{ij}/K_i$~\cite{comment3}. 
Suppose the
walker starts at node $i$ at time $t=0$, then the master equation for
the probability $P_{ij}$ to find the walker at node $j$ at time $t$ is
\begin{equation}\label{master_eq}
P_{ij}(t+1) = \sum_k  \frac{A_{kj}}{K_k} P_{ik} (t) \ .
\end{equation}
The largest eigenvalue of the corresponding time evolution operator is
$1$ corresponding to the stationary distribution
$P_j^{\infty}=\lim_{t\to\infty}P_{ij}(t)$, i.e.\ the infinite time 
limit~\cite{comment2}. An explicit expression for the transition
probability $P_{ij}(t)$ to go from node $i$ to node $j$ in $t$ steps
follows by iterating Eq.~(\ref{master_eq})
\begin{equation}
P_{ij}(t) = \sum_{j_1,\ldots,j_{t-1}} 
\frac{A_{i j_1}}{K_i} \cdot \frac{A_{j_1 j_2}}{K_{j_1}} 
\cdots  
\frac{A_{j_{t-1} j}}{K_{j_{t-1}}} \ .
\label{formal-solution}
\end{equation}
Comparing the expressions for $P_{ij}$ and $P_{ji}$ one sees
immediately that
\begin{equation}\label{Pij2Pji}
K_i P_{ij} (t) = K_j P_{ji} (t)  \ .
\end{equation}
This is a direct consequence of the undirectedness of the network. 
For the stationary solution, Eq.~(\ref{Pij2Pji}) implies that 
$K_i P^\infty_j = K_j P_i^\infty$, and therefore one obtains
\begin{equation}\label{stationary_solution}
P_i^\infty = \frac{K_i}{{\cal N}}
\end{equation}
with ${\cal N} = \sum_i K_i$. Note that the stationary distribution
is, up to normalization, equal to the degree of the node $i$ --- the
more links a node has to other nodes in the network, the more often it
will be visited by a random walker.

How fast is the random walk motion?
To answer to this question, we study the MFPT.
The first-passage probability $F_{ij}(t)$ from $i$ to $j$ after $t$ steps
satisfies the relation
\begin{equation}\label{PandF}
P_{ij}(t) = \delta_{t0} \delta_{ij} + \sum_{t'=0}^t
            P_{jj}(t-t')  F_{ij}(t') \ .
\end{equation}
The Kronecker delta symbol insures the initial condition $P_{ij}(0) = 
\delta_{ij}$~($F_{ij}(0)$ is set to zero). 
Introducing the Laplace transform $\tilde{f}(s) \equiv
\sum_{t=0}^\infty e^{-st} f(t)$, Eq.~(\ref{PandF}) becomes
$\widetilde{P}_{ij} (s) = \delta_{ij} + \widetilde{F}_{ij} (s) 
\widetilde{P}_{jj} (s)$,
and one has
\begin{equation}\label{LaplaceF}
\widetilde{F}_{ij} (s) = (\widetilde{P}_{ij}(s) - \delta_{ij}) / 
\widetilde{P}_{jj} (s) \ .
\end{equation}
In finite networks the random walk is recurrent~\cite{Hughes95}, 
so the MFPT is given by
$\langle T_{ij}\rangle = \sum_{t=0}^{\infty} t F_{ij} (t) 
= -\widetilde{F}'_{ij}(0)$.

Since all moments 
$R^{(n)}_{ij}\equiv \sum_{t=0}^{\infty} t^n ~ \{P_{ij}(t)-P_j^\infty\}$
of the exponentially decaying relaxation part of $P_{ij}(t)$ are
finite, one can expand
$\widetilde{P}_{ij}$ as a series in $s$ as
\begin{equation}
\widetilde{P}_{ij}(s) = \frac{K_j} {{\cal N} (1-e^{-s})}
+ \sum_{n=0}^\infty (-1)^n R^{(n)}_{ij} \frac{s^n}{n!} \ .
\end{equation}
Inserting this series into Eq.~(\ref{LaplaceF})
and expanding it as a power series in $s$, we obtain that 
\begin{equation}\label{tau}
\langle T_{ij} \rangle =
\left\{ 
\begin{array}{cl}
\frac{{\cal N}}{K_j}, & \quad \mbox{ for } j=i \\ [3mm]
\frac{{\cal N}}{K_j} \left[ R^{(0)}_{jj}- 
R^{(0)}_{ij} \right], & \quad \mbox{ for } j\neq i \end{array} \right. \ .
\end{equation}
A similar expression is derived in Ref.~\cite{Hughes95} for the MFPT of
the random walk in {\it periodic} lattices.

It is very interesting to note that the average return time
$\langle T_{ii}\rangle$ does not depend on the details of the 
global structure of
the network.  It is determined only by the total number of links and
the degree of the node. Since it is inversely proportional to the
degree, the heterogeneity in connectivity is well reflected in this
quantity.  In a SF network with degree distribution $P(K) \sim
K^{-\gamma}$, the MFPT to the origin $T_o$ also follows a power-law
distribution $P(T_o) \sim T_o^{-(2-\gamma)}$. The MFPT to the
origin distributes uniformly in the special case with $\gamma=2$.

Random walk motions between two nodes are asymmetric. The difference 
between $\langle T_{ij}\rangle$ and $\langle T_{ji} \rangle$ 
for $i\neq j$ can be written as (using Eq.~(\ref{tau}))
$$
\langle T_{ij}\rangle - \langle T_{ji} \rangle = 
{\cal N} \left( \frac{R^{(0)}_{jj}}{K_j}-\frac{R^{(0)}_{ii}}{K_i} \right) -
{\cal N} \left( \frac{R_{ij}^{(0)}}{K_j} - \frac{R_{ji}^{(0)}}{K_i} \right)  ,
$$
where the last term vanishes due to Eq.~(\ref{Pij2Pji}).
Therefore we obtain
\begin{equation}\label{tau_diff}
\langle T_{ij}\rangle - \langle T_{ji} \rangle =  C_j^{-1} - C_i^{-1},
\end{equation}
where $C_i$ is defined as
\begin{equation}\label{centrality}
C_i \equiv \frac{P_i^\infty}{\tau_i}\;,
\end{equation}
where $P_i^\infty=K_i/{\cal N}$ and
the characteristic relaxation time $\tau_i$ of the node $i$ is given by
\begin{equation}
\tau_i=R_{ii}^{(0)}=\sum_{t=0}^\infty \{ P_{ii}(t)-P_i^\infty\} \ .
\end{equation}
We call $C_i$ the {\em random walk centrality} since it quantifies how
central a node $i$ is located regarding its potential to receive
informations randomly diffusing over the network.  To be more precise:
Consider two nodes $i$ and $j$ with $C_i>C_j$. Assume that each of
them launches a signal simultaneously, which is wandering over the
network. Based on Eq.~(\ref{tau_diff}), one expects that the node with
larger RWC will receive the signal emitted by its partner
earlier. Hence, the RWC can be regarded as a measure for effectiveness
in communication between nodes.  In a homogeneous network with
translational symmetry, all nodes have the same value of the RWC. On
the other hand, in a heterogeneous network the RWC has a distribution,
which leads to the asymmetry in the random dynamic process.

The RWC is determined by the degree $K$ and $\tau$. The order of
magnitude of the characteristic relaxation time $\tau$ is related to
the second largest eigenvalue ({\em nota bene} \cite{comment2}) of the time
evolution operator in (\ref{master_eq}):
$P_{ii}(t)=P_i^\infty+\sum_{\lambda=2}^N a_i^{(\lambda)}
b_i^{(\lambda)}\Lambda_\lambda^t$, where $a^{(\lambda)}$ and $b^{(\lambda)}$
are the left and right eigenvectors, respectively, of the time
evolution operator belonging to the eigenvalue $\Lambda_\lambda$. If
we order the eigenvalues according to the modulus
($|\Lambda_2| \geq |\Lambda_3|\geq \ldots\geq |\Lambda_N|$) the 
asymptotic behavior is
$P_{ii}(t)-P_i^\infty\sim a_i^{(2)}b_i^{(2)}\Lambda_2^t$ and $\tau_i\approx
a_i^{(2)}b_i^{(2)}/|\ln|\Lambda_2||$. Thus the relaxation time $\tau_i$
has a node dependence only through the weight factor, which is
presumably weak. On the other hand, the degree dependence is
explicit.

\begin{figure}
\includegraphics*[width=\columnwidth]{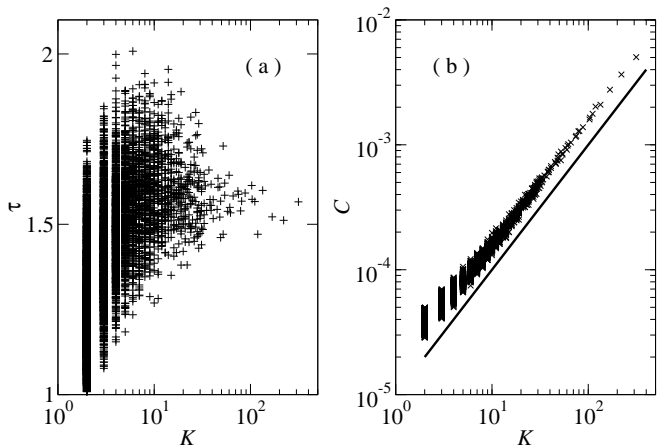}
\caption{(a) $\tau$ vs. $K$ and (b) $C$ vs. $K$ calculated in 
the Barab\'asi-Albert network with $N=10000$ and $m=2$.
The straight line in (b) has the slope 1.}\label{fig1}
\end{figure}

We examined the distribution of the RWC in the Barab\'asi-Albert~(BA)
network~\cite{Barabasi.Albert99}. This is a model for a growing SF
network; at each time step, a new node is added creating $m$ links
with other nodes which are selected with the probability proportional
to their degree. We grew the network, solved the master equation
numerically with the initial condition $P_i(t=0) = \delta_{ik}$, and
calculated the relaxation time $\tau_k$ for each $k$.
Figure~\ref{fig1}~(a) shows the plot of $\tau$ vs. $K$ in the BA
network of $N=10^4$ nodes grown with the parameter $m=2$.  The degree
is distributed broadly over the range $2\leq K \lesssim 400$. On the
other hand, the relaxation time turns out to be distributed very
narrowly within the range $1\lesssim \tau\lesssim 2$. We also studied
BA networks of different sizes, but did not find any significant
broadening of the distribution of $\tau$. So the RWC distribution
is mainly determined by the degree distribution.  In
Fig.~\ref{fig1}~(b) we show the plot of $C$ vs. $K$ in the same BA
network.  It shows that the RWC is roughly proportional to the degree.
Note however that the RWC is not increasing monotonically with the
degree due to the fluctuation of $\tau$ as seen in
Fig.~\ref{fig1}~(a).

The RWC is useful when one compares the random walk motions between
two nodes, e.g., $i$ and $j$ with $C_i > C_j$. On average a random
walker starting at $j$ arrives at $i$ before another walker starting
at $i$ arrives at $j$. Now consider an intermediate node $k$, which
may be visited by both random walkers. Since $\langle T_{ij}\rangle >
\langle T_{ji}\rangle$, it is likely that a random walker starting at
node $k$ will arrive at node $i$ earlier than at node $j$.  Although
this argument is not exact since we neglected the time spent on the
journey to the intermediate node, it indicates that nodes with larger
RWC may be typically visited earlier than nodes with smaller RWC by
the random walker.  If we interpret the random walker as an
information messenger, nodes with larger RWC are more efficient in
receiving information than nodes with smaller RWC.

\begin{figure}[t]
\includegraphics*[width=\columnwidth]{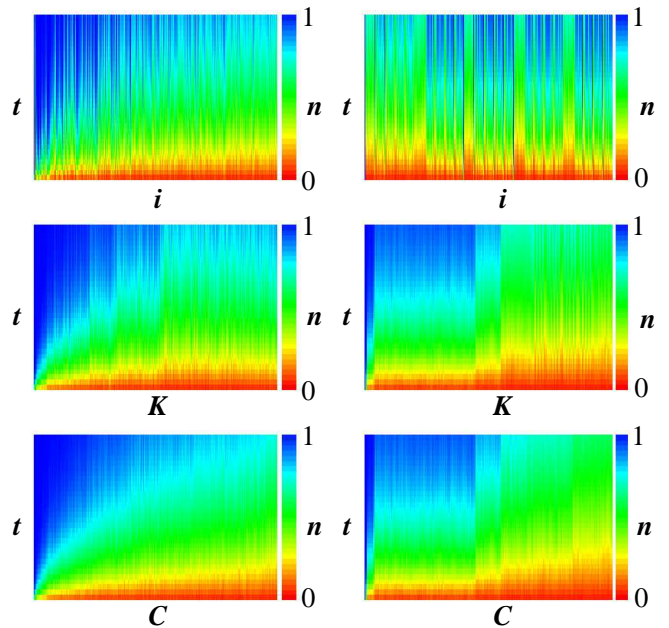}
\caption{(Color online) Time evolution of the fraction of walkers 
$n$ that pass through a node as a function of (from top to bottom) the node
index $i$, the node degree $K$ and the RWC $C$ of the BA 
network~(left column) and the hierarchical network~(right column). 
The value of $n$ at each time $t$ is represented in the gray scale/color 
code depicted at the right border of each plot.}
\label{fig2}
\end{figure}

We performed numerical simulations to study the relation between the
RWC and this efficiency. To quantify it, we consider a situation where
initially all nodes in a network are occupied by different random
walkers. They start to move at time $t=0$, and we measure $n_i$, the
fraction of walkers which have passed through the node $i$, as a
function of time $t$. It is assumed that the walkers do not interact
with each other. They may be regarded as a messenger delivering an
information to each node it visits. Then, with the information
distribution uniformly initially, $n_i$ is proportional to the amount
of information acquired by each node.  The argument in the previous
paragraph suggests that typically nodes with larger values of RWC have
larger value of $n_i$ at any given time.

The BA network~\cite{Barabasi.Albert99}
and the hierarchical network of Ravasz and Barab\'asi~\cite{cite:hnet} were
considered in the simulations.
The hierarchical network is a deterministic network growing via iteration; 
at each iteration the network is multiplied by a factor $M$. The emergent
network is scale-free when $M\geq 3$. Since it is a deterministic network,
several structural properties are known exactly~\cite{Noh03}. 
We measured $n_i$ in the BA network with $m=2$ and $N=512$ nodes 
and in the hierarchical network with $M=5$ and $N=5^4$ nodes
for $0\leq t\leq 2048$, which are presented in the left and the right
column of Fig.~\ref{fig2}, respectively. 
The value of $n_i$ is color-coded according to the reference 
shown in Fig.~\ref{fig2}. 

The time evolution of $n_i$ is presented in three different ways.  In
the first row, the nodes are arranged in ascending order of the node
index $i$. In the BA network, the node index corresponds to the time
step at which the node is added to the network. The indexing scheme
for the hierarchical network is explained in Ref.~\cite{Noh03}. In the
second row, the nodes are arranged in descending order of the degree
$K$ and in the third row they are arranged in descending order of the
RWC $C$.  At a given time $t$, the plot in the first row shows that
$n$ is non-monotonous and very irregular as a function of the node
index. As a function of the degree it becomes smooth, but still
non-monotonic tendencies remain. However, as a function of the RWC, it
becomes much smoother and almost monotonous.  
We calculated for each node $i$ the time $\tau'_i$ at which $n_i$ 
becomes greater than $1/2$. In the BA network, among all node pairs $(i,j)$
satisfying $\tau'_i<\tau'_j$, only $3\%$ violate the relation
$C_i>C_j$, whereas the number of pairs that violate the relation $K_i>K_j$
is five times larger.

In summary, we studied the random walk processes in complex
networks. We derive an exact expression for the mean first passage
time~(see Eq.~(\ref{tau})). The MFPT's between two nodes differ for
the two directions in general heterogeneous networks. We have shown
that this difference is determined by the random walk centrality $C$
defined in Eq.~(\ref{centrality}). Among random walk motions between
two nodes, the walk to the node with larger value of $C$ is faster
than the other.  Furthermore, it is argued that in a given time
interval nodes with larger values of $C$ are visited by more random
walkers which were distributed uniformly initially. We confirmed this
by numerical simulations on the BA and the hierarchical network. One
may regard the random walkers as informations diffusing through the
network.  Our results imply that information does not distribute
uniformly in heterogeneous networks; the information is centralized to
nodes with larger values of $C$. The nodes with high values of $C$
have the advantage of being aware of new information earlier than
other nodes.  On the other hand, it also implies that such nodes are
heavily loaded within an information distribution or transport
process. If the network has a finite capacity, the heavily loaded
nodes may cause congestions~\cite{Holme03}. Therefore much care should
be taken of the nodes with high $C$ values in network management.  In
the current work, we consider the random walks on undirected
networks.  The generalization to directed 
networks would be interesting.  And in order to study congestion, the
random walk motions with many {\em interacting} random walkers would
also be interesting. We leave such generalizations to a future work.

Acknowledgement: This work was supported by the Deutsche
Forschungsgemeinschaft (DFG) and by the European
Community's Human Potential Programme under contract 
HPRN-CT-2002-00307, DYGLAGEMEM.


\begin{references}
\bibitem{comment1} In this definition, all links are assumed to be
         equivalent. Evolution of the shortest path in weighted networks 
         is discussed in Ref.~\cite{Noh.Rieger02}.
\bibitem{comment3} In weighted networks, the hopping probability 
        may be written as $\widetilde{A}_{ij}/\widetilde{K_i}$ where
        $\widetilde{A}_{ij} = w_{ij}A_{ij}$ and $\widetilde{K}_i = \sum_j
        \widetilde{A}_{ij}$ with  a weight $w_{ij}>0$. All results in this 
        paper remain valid as long as the weight is 
        symmetric, i.e., $\widetilde{A}_{ij} = 
        \widetilde{A}_{ji}$.
\bibitem{comment2} The limit exists if and only if the network contains
        an odd loop. Then all other eigenvalues of the time evolution 
        operator satisfy $|\Lambda|<1$, otherwise 
        there exists an eigenvalue $\Lambda=-1$, for which the infinite 
        time limit does not exist. In such cases, one may redefine the
        RW model setting $A_{ii}=1$ to make the limit exist.
\end{references}
\end{document}